\documentclass{aa}
\usepackage{amssymb}
\usepackage{amsmath}
\usepackage{txfonts}
\usepackage{graphicx}
\usepackage{natbib}
\usepackage{rotating}
\usepackage{url}
 
\def\int{\textsl{INTEGRAL}}
\def\intsp{\textsl{INTEGRAL }}
\def\xte{\textsl{RXTE}}
\def\xtesp{\textsl{RXTE }}
\def\grs{GRS~1758\ensuremath{-}258}
\def\grssp{GRS~1758\ensuremath{-}258 }
\def\degsp{\ensuremath{^\circ} }
\def\deg{\ensuremath{^\circ}}

\begin{document}

\title{\textsl{INTEGRAL} and \textsl{RXTE} monitoring of
  GRS~1758$-$258\\ in 2003 and 2004}

\subtitle{A transition from the dim soft state to the hard state}

\author{Katja~Pottschmidt\inst{1}
\and Masha~Chernyakova\inst{2}\thanks{On leave from the Astro Space
  Center of the P.N.~Lebedev Physical Institute, 84/32 Profsoyuznaya
  Street, Moscow 117997, Russia.} 
\and Andrzej~A.~Zdziarski\inst{3}
\and Piotr~Lubi\'nski\inst{3,2}
\and David~M.~Smith\inst{4}
\and Nathan~Bezayiff\inst{4}}
\institute{Center for Astrophysics and Space Sciences, University of
California, San Diego, La Jolla, CA 92093-0424, USA 
\and \textsl{INTEGRAL} Science Data Centre, Chemin d'\'Ecogia
16, 1290 Versoix, Switzerland  
\and Nicolaus Copernicus Astronomical Center, Bartycka 18, 00-716
Warszawa, Poland 
\and Department of Physics, University of California, Santa Cruz,
Santa Cruz, CA 95064, USA} 

\offprints{kpottschmidt@ucsd.edu}
\titlerunning{Monitoring of GRS~1758$-$258}
\authorrunning{K.~Pottschmidt et al.}
\date{Received 20 August 2005 / Accepted 4 February 2006}
 
\abstract{The Galactic Center black hole candidate (BHC)
\object{GRS~1758$-$258} has been observed extensively within \int's
Galactic Center Deep Exposure (GCDE) program in 2003 and 2004, while
also being monitored with \xte. We present quasi-simultaneous PCA,
ISGRI, and SPI spectra from four GCDE observation epochs as well as
the evolution of energy-resolved PCA and ISGRI light curves on time
scales of days to months. We find that during the first epoch \grssp
displayed another of its peculiar dim soft states like the one
observed in 2001, increasing the number of observed occurrences of
this state to three. During the other epochs the source was in the
hard state.  The hard X-ray emission component in the epoch-summed
spectra can be well described either by phenomenological models,
namely a cutoff power law in the hard state and a pure power law in
the dim soft state, or by thermal Comptonization models. A soft
thermal component is clearly present in the dim soft state and might
also contribute to the softer hard state spectra. We argue that in the
recently emerging picture of the hardness-intensity evolution of black
hole transient outbursts in which hard and soft states are observed to
occur in a large overlapping range of luminosities (hysteresis), the
dim soft state is not peculiar. As noted before for the 2001 dim soft
state, these episodes seem to be triggered by a sudden decrease
(within days) of the hard emission, with the soft spectral component
decaying on a longer time scale (weeks). We discuss this behavior as
well as additional flux changes observed in the light curves in terms
of the existence of two accretion flows characterized by different
accretion time scales, the model previously suggested for the 2001
episode.

}

\maketitle

\section{Introduction}\label{sec:intro}

The hard X-ray source \grssp was discovered in 1990
\citep{mandrou:90a,sunyaev:91a} during observations of the Galactic
Center region performed with the \textsl{Granat} satellite. Most of
the time the source displays X-ray properties similar to the canonical
hard state of Galactic black hole binaries, i.e., a comparatively hard
spectrum with power law indices of $\Gamma$=1.4--1.9 and an
exponential cutoff above 100\,keV
\citep{kuznetsov:99a,main:99a,lin:00a} as well as strong short term
variability on frequencies up to 10\,Hz which can be characterized by
a flat-topped power spectrum \citep{smith:97a,lin:00a}. Based on these
X-ray properties and on the detection of a radio counterpart \citep[a
point source plus a double-sided jet structure,][]{rodriguez:92a}
\grssp is considered to be a microquasar.

With the exception of rare dim soft states that can last up to several
months, the X-ray emission of \grssp is persistent. In contrast to the
canonical picture for persistent black hole binaries, however, \grssp
most likely has a low mass binary companion and is accreting via Roche
lobe overflow. Three faint IR counterpart candidates have been
identified recently, multi-band photometry and near-infrared
spectroscopy characterizing the brightest of them as a probable K0~III
giant and the other two as main sequence A stars
\citep{marti:98a,goldwurm:01a,eikenberry:01a,rothstein:02a,heindl:02b}.
Taking into account the 18.45$\pm$0.10\,day binary orbit
\citep{smith:02a}, the radius of a K giant is consistent with Roche
lobe overflow, while the other counterpart candidates are too small
\citep{rothstein:02a}.

From 1990 to 1998 the source was monitored in the hard X-ray regime
with \textsl{Granat}/SIGMA, establishing it as persistent hard state
black hole binary but also revealing a factor of eight variability in
the 40--150\,keV flux \citep{kuznetsov:99a}. At softer X-rays
monitoring of \grssp (and its ``sister source''
\object{1E~1740.7$-$2942}) with \xtesp started with monthly
observations in 1996 and is still on-going in 2005, with two
observations each week \citep[][this
work]{main:99a,smith:97a,smith:01a,smith:02a,smith:01b}. This campaign
led to the discovery that in \grssp (and 1E~1740.7$-$2942) the
observed spectral hardness is not anti-correlated with the 2--25\,keV
flux but rather correlated with the flux derivative -- in the sense
that the spectrum is softest when the 2--25\,keV count rate is
dropping. This behavior is different from the prototype hard state
black hole binary \object{Cyg~X-1} which is showing the canonical
anti-correlation of spectral hardness and soft X-ray flux. It has been
interpreted as an indication for the presence of two
\textsl{independent} accretion flows supplied with proportional
amounts of matter at large radii which are then accreted on different
time scales \citep{main:99a,smith:01b}: a hot, e.g., ADAF-type,
accretion flow, reacting quickly to changes in the accretion rate, and
a large accretion disk with a long viscous time scale (consistent with
accretion via Roche lobe overflow). In addition, \citet{lin:00a}
performed a radio to $\gamma$-ray multi-wavelength study of the hard
state as observed in 1997 August, including spectral modeling and high
time resolution analyses. Applying the thermal Comptonization model
\texttt{compTT} \citep{tit:94} they obtained a temperature of 52\,keV
and an optical depth of 3.4 for the hot plasma. \citet{sidoli:02a}
obtained very similar values using \texttt{compTT} to model a broad
band \textsl{BeppoSAX} spectrum of the source obtained in 1997 April.

A weak soft excess is sometimes seen in the hard state spectra of
\grssp \citep{heindl:98a,lin:00a} or cannot be excluded to be present
\citep{mereghetti:97a}. It has also been observed in conjunction with
a slightly reduced hard X-ray flux during an intermediate state in
1993 March/April \citep{mereghetti:94a}. A similar recent episode in
2000 September has been characterized as an intermediate state based
on the \xtesp monitoring observations \citep[][and references
therein]{heindl:02b}. Modeling an \textsl{XMM-Newton}/EPIC-MOS
spectrum obtained during this time, \citet{goldwurm:01a} found that in
addition to a comparatively soft power law ($\Gamma\sim$2) a blackbody
component ($kT_{\text {in}}\sim$320\,eV) is required.

On two occasions a much softer and dimmer state than the persistent
hard state with occasional softening has been observed: (i) A sudden
drop of the 2--25\,keV rate between one \xtesp monitoring observation
and the next occurred in 2001 February, with an estimated
\textsl{decrease} of $\sim$35\% of the 1.3--200\,keV luminosity within
$\sim$20\,d after the transition \citep{smith:01a}. An extreme case of
the unusual flux derivative/hardness relation described above, this
behavior is different from the canonical black hole state behavior
where the soft state corresponds to a higher accretion rate and a
higher bolometric luminosity. For the 1996 soft state of Cyg~X-1,
e.g., \citet{zdz:02a} find a 3--4 times higher bolometric luminosity
than for the typical hard state, an increase even higher than the
$\sim$50--70\% estimated previously \citep{zhang:97a}. In
Sect.~\ref{sec:discussion} we suggest that the dim soft state can be
better understood when compared to outbursts of BHC transients than to
(focused) wind accretors like Cyg X-1. \citet{smith:01a} also found
that the transition to the 2001 dim soft state of \grssp was mainly
due to a decreasing and softening power law component
($\Gamma\sim$2.75 in 2001 March), revealing a soft component
($kT_{\text {in}}\sim$460\,eV in 2001 March). As predicted by
\citet{smith:01c} based on the two-flow accretion model, the soft
component decayed more slowly than the hard one, on a time scale of
$\sim$28\,days. Displaying a complex structure, including a partial
recovery of the count rates in 2001 July, the dim soft state lasted
until 2001 December \citep[][see also this
work]{heindl:02b}. \textsl{Chandra}/ACIS-HETGS
\citep{heindl:02a,heindl:02b} and \textsl{XMM-Newton}/RGS
\citep{miller:02a} observations in 2001 March support the picture that
the decaying soft flux is emitted by an accretion disk. (ii) It can be
assumed that \textsl{Granat}/SIGMA exposures in fall 1991 and spring
1992 when the 40--150\,keV flux was below the detection limit
\citep{gilfanov:93a} found the source in a similar dim soft state
\citep{smith:01a,miller:02a}. This is supported by the analysis of a
1992 March \textsl{ROSAT}/PSPC spectrum by \citet{grebenev:97a}
resulting in a power law index of $\Gamma\sim$2.5.

In the following we report results of monitoring observations of
\grssp with \intsp and \xtesp in 2003 and 2004. While the source was
in its usual variable hard state during most of the time, the data
obtained in spring 2003 clearly correspond to another dim soft state,
although it did not progress as far as the 2001 dim soft state before
the hard X-ray emission recovered again. In
Sect.~\ref{sec:observations} of this paper we describe the observing
strategy of the two monitoring programs and explain the data
extractions performed to obtain broad band PCA-ISGRI-SPI spectra and
multi-band PCA and ISGRI light curves. In Sect.~\ref{sec:evol} the
long term light curves and flux changes are described and in
Sect.~\ref{sec:spectra} we present results of modeling the broad band
spectra with phenomenological and thermal Comptonization models. The
results, especially the observation of another weak soft state, are
discussed in the light of current black hole outburst and accretion
models in Sect.~\ref{sec:discussion}. Our conclusions are summarized
in Sect.~\ref{sec:conclusions}.

\section{Observations and Data Reduction}\label{sec:observations}

During 2003 and 2004 the guaranteed time program amounted to 30--35\%
of \int's observing time and was mainly dedicated to the
Galactic Plane Scan (GPS) and Galactic Center Deep Exposure (GCDE)
projects within the \intsp team's Core Programme. Especially the
GCDE\footnote{Concentrating on 5\degsp around the Galactic Center but
with pointing positions reaching out to Galactic longitudes and
latitudes of about $\pm30$\degsp and $\pm20$\deg, respectively, from
it.} provided a wealth of data on \grs. All Core Programme data up to
January 2005 as well as all data of the source public at that time
have been included into the analysis presented here. Our \intsp data
are grouped into four data sets observed in spring and fall of 2003
and 2004 which in the following shall be called epoch~1--4. See
Table~\ref{tab:obslog} for more details.

\begin{table}
 \caption{ \intsp observing epochs for \grs, giving the exposure times
  of the summed spectra analyzed for each epoch and instrument,
  including the \xte/PCA.}\label{tab:obslog}
\begin{tabular}{ccrrr}
\hline
Epoch$^a$& Start \& End Date & ISGRI & SPI & PCA$^{b,c}$\\
         &                   & [ks]  & [ks]& [ks]\\
\hline
1 & 2003/02/28--2003/04/23    &  511&     -- & 27\\
2 & 2003/08/19--2003/10/14    & 1889&   1358 & 19\\
  &                           &   & 1141$^d$ &   \\ 
3 & 2004/02/17--2004/04/20    &  578&    759 & 23\\
4 & 2004/08/21--2004/10/28    &  467&    615 & 10\\
\hline
\end{tabular}

\vspace{2mm}

$^a$The four epochs correspond to \intsp orbits 46--66, 103--122,
164--185, and 226--249.\\  
$^b$The PCA exposure includes only data from PCU~2, see text
for more details.\\
$^c$Data from the following \xtesp programs are included: 50107,
80102, and 90102.\\   
$^d$SPI spectra for two non-overlapping data sets within this epoch
have been extracted (orbits 103--111 and 112--122), see text for more details.

\end{table}

\begin{figure}
\includegraphics[width=88mm]{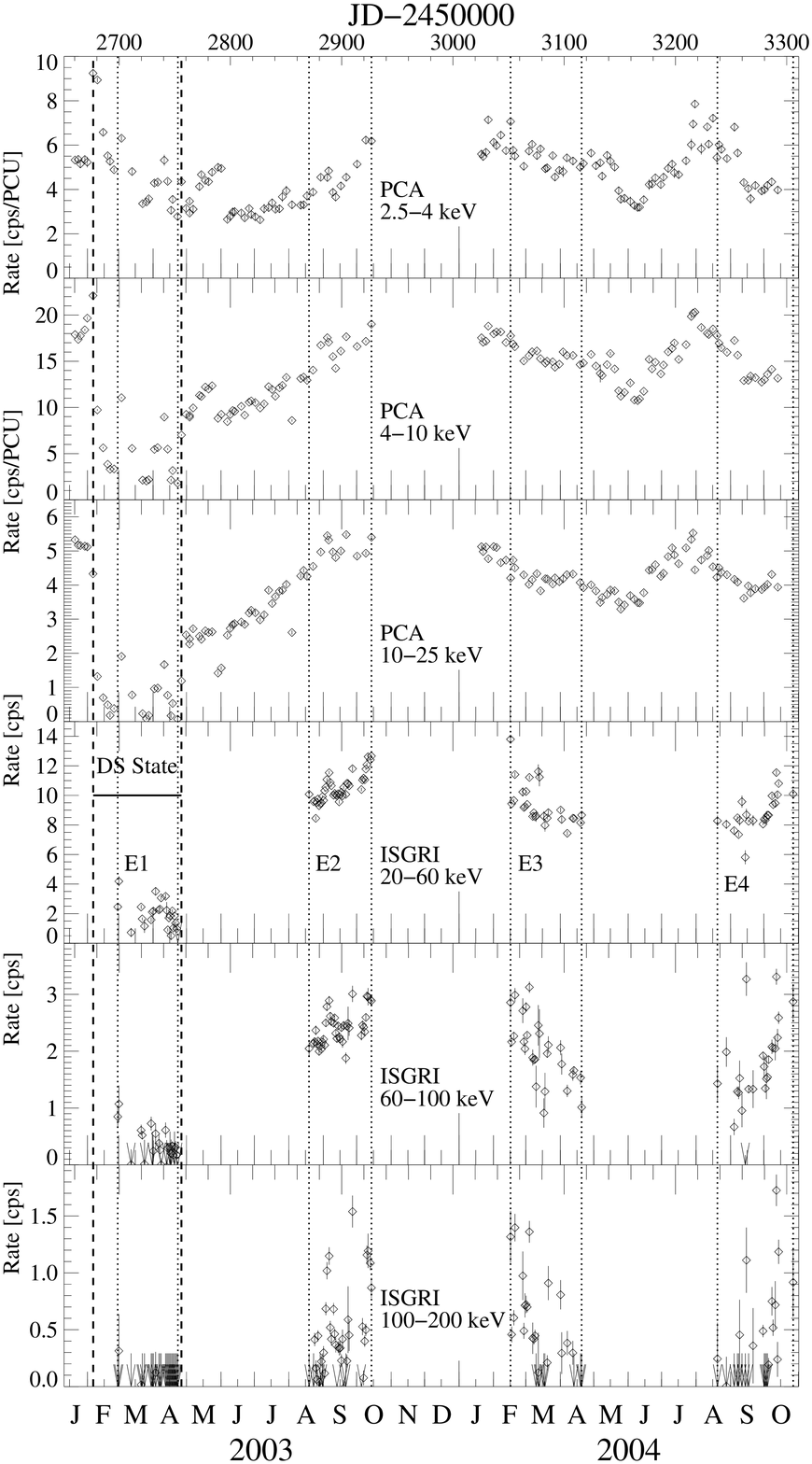}
\caption{\int/ISGRI and \xte/PCA light curves of \grssp in the
  indicated energy ranges as observed in 2003 and 2004. The ISGRI
  light curves have been rebinned to a resolution of 1\,d. Arrows
  indicate non-detections. For the PCA the average count rate of each
  monitoring observation is shown in all energy bands. Vertical dashed
  and dotted lines denote the dim soft state (``DS State'') and the
  four \intsp observation epochs (``E1''--``E4''), respectively.}
\label{fig:lcpanel}
\end{figure}

We used version 4.2 of the Offline Scientific Analysis package for
\intsp to extract spectra and light curves of \grssp obtained by the
\intsp Soft Gamma Ray Imager \citep[ISGRI;][]{lebrun:03a} detector as
well as spectra from the SPectrometer on \intsp
\citep[SPI;][]{vedrenne:03a}\footnote{At the time of submission of
this paper, OSA~5.0 had just become available. The conclusions
presented in this work do not depend on the OSA version. Since OSA~4.2
extractions of ISGRI spectra for sources as faint as \grssp require
the use of the special procedure described in this section and since
no experience with using OSA~5.0 for faint sources existed prior to
submission, we decided to continue using OSA~4.2.}. See
\url{http://integral.esac.esa.int/workshops/Jan2005/session1/lubinski_cross.pdf}
for information on instrument cross-calibration issues for OSA
4.2. Due to the grid nature of the observations and the usually hard
source spectrum, the smaller field of view Joint European X-ray
Monitor \citep[JEM-X;][]{lund:03a} did not yield any useful data
covering the epoch time scales. In order to extract the ISGRI data
products, all pointings (``science windows'', ``ScWs'') in which the
offset of the source from the spacecraft pointing direction was
smaller than 10\degsp have been taken into account. For offsets in
this range systematic effects in the Crab calibration spectra show the
same general trends as for pointings in the fully coded field of
view. This selection amounts to $\sim$1920 ScWs with an exposure of
approximately 1800\,s each. Average spectra for the four epochs were
built by producing images in 12 energy bands for each ScW in a given
epoch, extracting the \grssp source flux from each image, and
averaging the source fluxes of all ScWs in a given energy band using
standard weighting techniques. This method is described in the ISGRI
user manual
(\url{http://isdc.unige.ch/doc/tec/um/user_support/ibis/ibis_4.2.pdf})
and is the recommended procedure for all but the brightest
sources. For the coded aperture instrument ISGRI the diffuse Galactic
background is part of the background removed when reconstructing the
sky images out of detector shadowgrams
\citep{goldwurm:03a,terrier:03a}. For the spectral modeling we use the
ancillary response file ``isgr\_arf\_rsp\_0006.fits'' and a rebinned
version of the response matrix ``isgr\_rmf\_grp\_0012.fits''
distributed with OSA~4.2. In addition, light curves with a time
resolution of 1000\,s were produced in three energy bands: 20--60,
60--100, and 100--200\,keV.

During the first epoch the source was too soft to be detected by the
SPI instrument. For the remaining three epochs the same ScWs as for
ISGRI were used to produce epoch-summed SPI spectra, with the
exception of epoch 2, where successful OSA runs could only be obtained
by splitting the SPI data into two subsets. The difference in the
exposure times given for ISGRI and SPI in Table~\ref{tab:obslog} are
mainly due to ISGRI's dead-time. The SPI spectra were extracted over
an energy range of 20--500\,keV (25 bins) using the SPIROS package
within OSA, applying maximum likelihood optimization statistics
\citep{skinner:03a}. We set the number of pseudo detectors to 84
(i.e., including events located near borders between the physical SPI
detectors and registered in more than one of them) and selected
background correction method 5 (detector averaged count rate
modulation model). The input catalog of sources consisted of the 18
sources seen in the ISGRI 20--60\,keV mosaic images. Alternative
parameter settings were tested, like changing the number of pseudo
detectors to 18 (i.e., including only single events), using background
model 2 (each detector scaled separately), or allowing sources to be
variable (SEL\_FLAG=2). None of these alternatives lead to a
significant change in the obtained count rates. Applying an
alternative extraction method optimized for recovering spatially
extended emission, \citet{strong:03a} find that the diffuse Galactic
background spectrum is of roughly power law shape, falling with a
slope of 2.5--3. We verified that adding such a component to only the
SPI data does not change the best fit parameters of the
multi-instrument fits discussed in section~\ref{sec:spectra} and that
the normalization of the new power law is consistent with zero. Note
that according to a study by \citet{dubath:05a} based on observations
and simulations, SPI fluxes of sources \textsl{with known positions}
can be well recovered down to a source separation of at least
0$\fdg{}$5. With an angular distance of 0$\fdg{}$66 from \grssp the
nearest source, the bright LMXB \object{GX~5$-$1}, should therefore
not contaminate our \grssp SPI spectra. For the data sets presented
here a careful inspection of the spectra of both sources shows no
indication of contamination with the possible exception of one bin
affected by an overcorrection (undercorrection) of the 66.7\,keV
background line for \grssp (GX~5$-$1). We find, however, that
excluding this bin from the analysis does not change the results.

In order to characterize the source behavior at softer X-ray energies
we use data from \xte's Proportional Counter Array
\citep[PCA;][]{jahoda:96} obtained during the on-going \xtesp
monitoring campaign. Under this program 1.5\,ks snapshots of \grssp
have been taken monthly in 1996, weekly from 1997 through 2000, and
approximately twice each week since then. For a description of the
offset observing strategy applied to avoid nearby sources and of the
extra background measurements taken to minimize the influence of the
diffuse Galactic emission see \citet{smith:97a} and
\citet{main:99a}. This procedure has been successfully evaluated using
data from the pronounced dim soft state in 2001.  Reduction of the PCA
data was performed using the HEASOFT package version~5.3.1. The
responses were generated using pcarsp version~10.1 (see
\url{http://lheawww.gsfc.nasa.gov/docs/xray/xte/pca/} for more
information on the \xte/PCA energy calibration under this HEASOFT
version). Average spectra for the four \intsp observing epochs were
produced. In addition, long term light curves consisting of the
average count rates of each PCA monitoring pointing were generated in
three energy bands (2.5--4, 4--10, and 10--25\,keV) and for the
overall 2.5--25\,keV light curve. We only use data from PCA's top
layer to optimize the signal to noise ratio. For the average spectra
we additionally decided to select data from one of PCA's five
Proportional Counter Units (PCUs), namely from PCU~2,
only\footnote{PCU~2 and PCU~3 are the units mainly used for
monitoring. In the data sets discussed here PCU~1 and PCU~4 only
contain $\sim$15--30\% of the exposure of PCU~2 or PCU~3.}. The loss
of additional PCA exposure is acceptable since our aim is to study the
broad band spectral continuum (and not, e.g., to perform deeper iron
line studies) with emphasis on characterizing the hard spectral
component, i.e., on the \intsp data. This strategy also allows us to
further minimize systematic effects due to PCU cross-calibration. See
Table~\ref{tab:obslog} for the total exposure times of the
epoch-summed PCA spectra. Note that the All Sky Monitor (ASM) on
\xtesp is not well suited to observe \grs: the source's daily
1.3--12.2\,keV ASM rate averaged over the time of the \intsp mission
up 2005 March, e.g., is 2.0$\pm$2.5\,cps. Also, since the absorbed
soft flux does not change much in the dim soft state (see next
section), no change is seen in the ASM light curve around epoch~1.

\section{Data Analysis and Results}\label{sec:analysis}

\begin{figure}
\includegraphics[width=88mm]{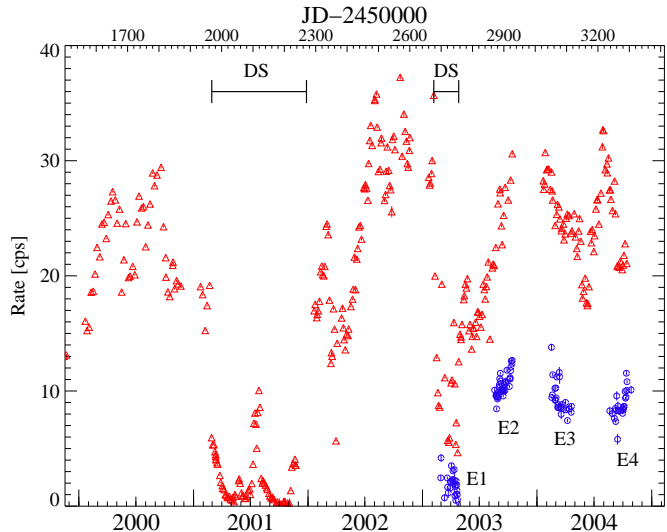}
\caption{Evolution of the 2.5--25\,keV PCA count rates of \grssp since
  2000 (open triangles) and of the 20--60\,keV ISGRI count rates in
  2003/2004 (open circles, see also Fig.~\ref{fig:lcpanel}). The
  binning is the same as in Fig.~\ref{fig:lcpanel}. The dim soft
  states are denoted by ``DS'' and the four \intsp observation epochs
  are again labeled ``E1''--``E4''.}
\label{fig:longlc}
\end{figure}

\subsection{Long Term Light Curves and Flux Changes}\label{sec:evol}

Fig.~\ref{fig:lcpanel} shows the evolution of the \intsp and \xtesp
light curves of \grssp during the monitoring in 2003 and 2004 in
several energy bands, spanning a total energy range of
2.5--200\,keV. The \intsp light curves have been rebinned to a
resolution of one day, for \xtesp the average count rate of each PCA
data set is plotted, normalized to one PCU. The count rates during
epoch~1 are significantly lower in all energy bands above
4\,keV. Above 100\,keV the source is not detected in epoch~1. The PCA
measurements during and between epochs~1 and 2 suggest that the former
almost exactly covers the last two months of a $\sim$3 months long
period during which the source was in a dim soft state and that a
transition to the more common hard started with the end of
epoch~1. 

This picture is supported by the average flux values determined from
our spectral fits (Table~\ref{tab:fluxes}). The 4--100\,keV flux is
considerably reduced in epoch~1. Consistent with the light curves the
\textsl{absorbed} 2.5--4\,keV flux does not change much between the
soft and hard state epochs. The \textsl{unabsorbed} flux extrapolated
to slightly lower energies, namely 2--4\,keV, reveals an overall
brightening at very low energies in epoch~1, however. A similar
behavior was also observed during the onset of the 2001 dim soft state
\citep{smith:01a}. Different from 2001, though, the 2003 dim soft
state starts with a short peak in the soft 2.5--4\,keV light curve,
coinciding with the decay above 10\,keV.  The 4--10\,keV light curve
shows a superposition of both trends, with the flare dominating first
and then the decay.

\begin{table}
\caption{Average model flux for each epoch based on the best fit values
of Table~\ref{tab:cutoffpl}.}\label{tab:fluxes}
\begin{tabular}{lrrrr}
\hline
\hline
                  &         Epoch 1&      Epoch 2&      Epoch 3&      Epoch 4\\
\hline
$F_{2.5-4}^b$     &             1.6&          1.4&          1.4&          1.5\\
$F_{2-4}^a$       &             4.2&          2.3&          2.5&          2.8\\
$F_{3-9}^{a,c}$   &             1.9&          4.7&          4.5&          4.7\\
$F_{4-100}^a$     &             2.5&         22.9&         18.7&         18.6\\
\hline
\end{tabular}

$^a$ Unabsorbed flux in units of $10^{-10}$ erg cm$^{-2}$
s$^{-1}$.\\
\hspace*{0.15cm} The $N_{\text H}$ values of Table~\ref{tab:cutoffpl} have
been used for the flux correction. \\ 
$^b$ Absorbed flux in units of $10^{-10}$ erg cm$^{-2}$ s$^{-1}$.\\
$^c$ See comparison with flux changes of \object{GX~339$-$4} in
Sect.~\ref{sec:discussion}. 
\end{table}

To put the \intsp observing epochs into the broader context of the
source history, Fig.~\ref{fig:longlc} shows the 2.5--25\,keV light
curve from the PCA monitoring since 2000 as well as the 20--60\,keV
ISGRI light curve again. The dim soft state in 2001 is readily
apparent, including two instances within the off phase (2001 May and
2001 July/August) where the source partly turns on again. The
2.5--4\,keV and 10--25\,keV overview light curves shown by
\citet{heindl:02b} include these turn-ons. In their Fig.~1 it can be
seen that the soft emission only reaches its minimum after the second
turn-on. The soft emission decays slower than the hard emission after
each turn-on, consistent with the two-flow scenario. The decrease of
the 2.5--25\,keV rate in 2003 February is slower than the rapid
initial drop in 2001 February, but the 10--25\,keV light curve in
Fig.~\ref{fig:lcpanel} reveals that the hard emission decreases on a
similar time scale as in 2001, i.e., within about a week. The 2003 dim
soft state is shorter and not declining below $\sim$4\,cps/PCU in the
2.5--25\,keV band, though. In addition to the dim soft states there is
considerable long term variability present in the light curves, the
\intsp 20--60\,keV ISGRI rate, e.g., varies by a factor of 30--40\%
within each hard state epoch. Furthermore, the 2.5--25\,keV PCA light
curve shows several sudden count rate drops, less severe or shorter
than the dim soft states, e.g., in 2002 March or between epochs~3 and
4 or during epoch~4.

\subsection{Spectral Analysis}\label{sec:spectra}

\begin{figure*}
\centering
\includegraphics[width=\textwidth]{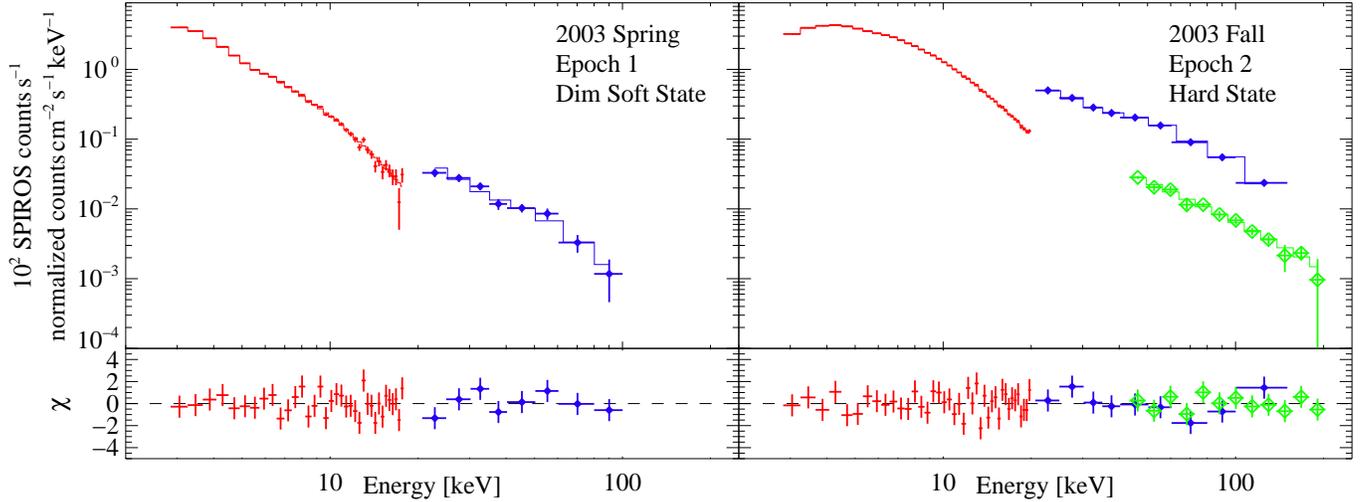}
\caption{Summed counts spectra for the \grssp monitoring observations
  of 2003 spring (epoch~1) and 2003 fall (epoch~2) with the best fit
  \texttt{compTT} model and the corresponding residuals. Small dots
  are PCA, filled diamonds ISGRI, and open diamonds SPI data. For
  reasons of clarity only the first of the two SPI spectra for epoch~2
  is shown but both data sets have been used to derive the plotted
  model and residuals as well as the best fit parameters listed in
  Tables~\ref{tab:cutoffpl} and~\ref{tab:comptt}. Note that ``count
  rates'' delivered by the SPI extraction software SPIROS are not
  directly comparable to those of other instruments and that here and
  in Fig.~\ref{fig:ldata34} the SPI spectra have additionally been
  multiplied by a factor of 100 for display
  purposes.}\label{fig:ldata12}
\end{figure*}

\begin{figure*}
\centering
\includegraphics[width=\textwidth]{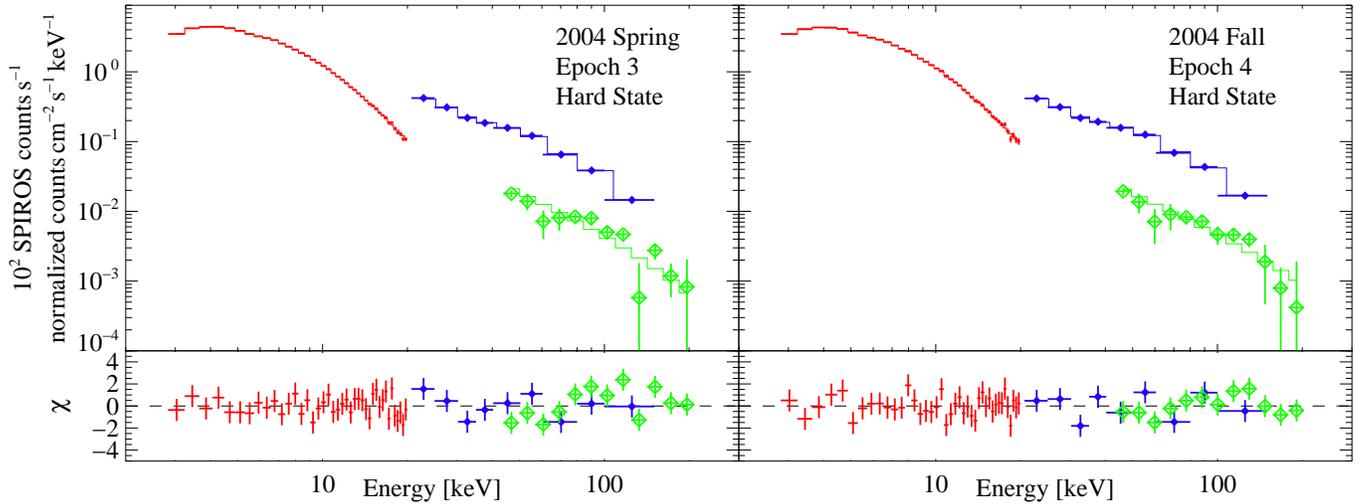}
\caption{The same as Fig.~\ref{fig:ldata12} but for the observations
  of 2004 spring (epoch~3) and 2004 fall
  (epoch~4).}\label{fig:ldata34}
\end{figure*}

\subsubsection{Models and Data Preparation}

Below 100\,keV \grssp has often been modeled by an absorbed power law
\citep{sunyaev:91a,mereghetti:97a,main:99a}. \citet{kuznetsov:99a}
found that an exponential cutoff power law fits the 1990--1997
\textsl{GRANAT} data above 100\,keV better than a simple power law and
\citet{lin:00a} obtained good fits to their joint \xte/PCA,
\xte/HEXTE, and \textsl{CGRO}/OSSE spectrum of 1997 with a cutoff
power law.  Thermal Comptonization has also been shown to provide a
good description of the hard state spectra
\citep{kuznetsov:99a,lin:00a,keck:01a}. As reported in
Sect.~\ref{sec:intro}, an additional weak thermal component can be
present in the hard state \citep{heindl:98a,lin:00a} which is more
clearly revealed in the intermediate
\citep{mereghetti:94a,goldwurm:01a} and especially the dim soft state
\citep{smith:01a,miller:02a,heindl:02b}.

From initial power law fits to our \intsp data alone, we found that
for epochs~2 to 4 a cutoff is required. This is imposed by the ISGRI
data sets. Due to SPI's comparatively small effective area, the SPI
data do not carry enough weight to further constrain the cutoff
energy. Our basic phenomenological model for the simultaneous fits to
the summed \int/\xtesp spectra of each epoch thus consists of an
absorbed cutoff power law plus a Gaussian Fe~K$\alpha$ line (see
Sect.~\ref{sec:fitting} for a discussion of the need to include the
line), with an additional multicolor disk blackbody component if
required. We also applied a thermal Comptonization model
\citep[\texttt{compTT};][]{tit:94} to all four epochs, again including
absorption, Fe emission, and the optional disk blackbody as well as
allowing for a reflected Comptonized component
\citep{magdziarz:95a}. Normalization differences between the
instruments are taken into account in all fits by a multiplicative
constant, set to 1 for the PCA. The exact model compositions of both,
the phenomenological and the physical model, can be found in the
captions of Tables~\ref{tab:cutoffpl} and \ref{tab:comptt} for all
epochs.

We used XSPEC version~11.3.1t to perform the fits. Consistent with the
recommendations of the calibration teams, systematic errors of 0.5\%
and 2\% had to be added to all PCA and ISGRI spectra,
respectively. PCA data from 3--20\,keV, ISGRI data from 20--150\,keV,
and SPI data from 40--200\,keV were taken into account in all fits,
with the exception of epoch~1 where PCA data up to only 18\,keV and
ISGRI data up to only 100\,keV were included. Both modeling approaches
resulted in good descriptions of the data and produced similar
$\chi^2_{\text {red}}$ values for given
epochs. Tables~\ref{tab:cutoffpl} and \ref{tab:comptt} list the best
fit parameters and $\chi^2$ values for the power law and the
Comptonization models, respectively. Single parameter uncertainties
are given on a 90\% confidence level. The results quoted for epoch~2
contain both SPI data sets. Without the 2.5\,Ms of SPI data, the
$\chi^2_{\text {red}}$ values of the epoch~2 fits are in better
agreement with the quality of the other fits, e.g., $\chi^2_{\text
{red}}$=1.1 for the epoch~2 \texttt{cutoffpl} fit (with no significant
changes of the best fit parameters). Fig.~\ref{fig:ldata12} and
Fig.~\ref{fig:ldata34} show the counts spectra, best fit models, and
residuals for the \texttt{compTT} fits.

Since the calibration of the \intsp instruments, especially ISGRI, is
work in progress, we expect that the best fit parameters
characterizing the hard spectrum will be refined and updated in future
iterations of this work. In this iteration we interpret them as
indicators for general trends (e.g., the state change, qualitative
consistency with canonical values, etc.). Modeling the spectra with
Comptonization models also taking non-thermal electron distributions
into account like \texttt{compPS} \citep{poutanen:96a} or
\texttt{eqpair} \citep{coppi:99a} will also be part of future
work. However, consistency checks have been performed, applying the
\texttt{compPS} model in a form comparable to our \texttt{compTT} fits
(thermal electrons, slab geometry, optional multicolor disk
blackbody). We obtain fits of similar quality, with seed photon
temperatures, plasma temperatures and optical depths consistent with
the \texttt{compTT} results. The reflection fraction obtained with
\texttt{compPS} is systematically higher, though, e.g., 24\% for
epoch~2 compared to 10\% with \texttt{compTT}. A similar trend of
lower reflection fractions obtained with \texttt{compTT} was also
observed between \texttt{eqpair} and \texttt{compTT} fits to Cyg~X-1
\int/\xtesp spectra \citep{pottschmidt:03a,pottschmidt:04a}.  While
the \texttt{compTT}/\texttt{eqpair} discrepancy is likely due to the
omission of relativistic smearing of the reflection spectrum in
\texttt{compTT}, as recently suggested by \citet{wilms:05a} on the
basis of \texttt{compTT} and \texttt{eqpair} fits to several hundred
\xtesp monitoring observations of Cyg~X-1, this is not the case here
since our \texttt{compPS} fits do not include relativistic smearing.

\begin{table}
\caption{Best fit parameters for the power law models. The full model
in XSPEC notation for epoch~1 is \texttt{const$\times$phabs
[diskbb+gauss+power]}, for epoch~2 it is
\texttt{const$\times$phabs[gauss+cutoffpl]} and for epochs~3 and 4
\texttt{const$\times$phabs[diskbb+gauss+cutoffpl]}. Parameters shown
are the hydrogen column density $N_{\text H}$, the inner accretion
disk temperature $kT_\text{in}$ and its normalization
$A_\text{disk}=((R_\text{in}/\text{km})/(D/10\,\text{kpc}))^2\cos i$,
the power law index $\Gamma$ and the power law cutoff energy
$E_{\text {cutoff}}$, the energy $E_\text{Fe}$ and equivalent width
$EW_\text{Fe}$ of the Gaussian Fe K$\alpha$ line, and the flux
normalization constants of the individual instruments with respect to
the PCA, $c_\text{ISGRI}$ and $c_\text{SPI}$.}\label{tab:cutoffpl}
\begin{tabular}{lrrrr}
\hline
\hline
                                    &                         Epoch 1&                       Epoch 2  &                                   Epoch 3&                            Epoch 4\\
\hline  
$N_{\text {H}}/10^{22}$ [cm$^{-2}$] &          1.66$^{+1.60}_{-0.41}$&          1.11$^{+0.25}_{-0.20}$&                                      1.50&             1.50$^{+3.18}_{-1.20}$\\
$kT_{\text {in}}$ [eV]              &               477$^{+11}_{-27}$&                              --&                        679$^{+145}_{-66}$&                536$^{+250}_{-222}$\\
$A_{\text {disk}}/10^{3}$           &             2.7$^{+0.9}_{-0.1}$&                              --&                    0.04$^{+0.05}_{-0.02}$&             0.16$^{+0.41}_{-0.12}$\\
$\Gamma$                            &          2.29$^{+0.10}_{-0.05}$&          1.54$^{+0.01}_{-0.02}$&                    1.59$^{+0.03}_{-0.04}$&             1.69$^{+0.05}_{-0.05}$\\
$E_{\text {cutoff}}$ [keV]          &                              --&               185$^{+22}_{-17}$&                         136$^{+13}_{-16}$&                  246$^{+26}_{-56}$\\
$E_{\text {Fe}}$ [keV]              &          6.40$^{+0.12}_{-0.19}$&          6.52$^{+0.21}_{-0.27}$&                    6.66$^{+0.15}_{-0.24}$&             6.57$^{+0.13}_{-0.37}$\\ 
$EW_{\text {Fe}}$ [eV]              &                           146.0&                            59.2&                                      53.8&                               49.0\\ 
$c_{\text {ISGRI}}$                 &          0.70$^{+0.06}_{-0.06}$&          0.85$^{+0.02}_{-0.02}$&                    0.83$^{+0.02}_{-0.03}$&             0.88$^{+0.01}_{-0.02}$\\
$c_{\text {SPI}}$                   &                              --&          0.99$^{+0.02}_{-0.04}$&                    1.00$^{+0.05}_{-0.10}$&             1.00$^{+0.11}_{-0.11}$\\
$\chi^2_{\text {no Fe}}/{\text {dof}}$&                       61.3/39&                         88.6/69&                                   83.3/55&                            60.2/55\\
$\chi^2/{\text {dof}}$              &                         37.7/35&                         55.7/65&                                   56.7/52&                            49.4/51\\
$\chi^2_{\text {red}}$              &                            1.08&                            0.86&                                      1.09&                               0.97\\
\hline
\end{tabular}

\end{table}

\begin{table}
\caption{Best fit parameters for the \texttt{compTT} model. The full
model in XSPEC notation is
\texttt{const$\times$phabs[diskbb+gauss+compTT+reflect(compTT)]},
where for epoch~2 the \texttt{diskbb} component and for epochs~1 and 4
the \texttt{reflect(compTT)} component turned out not to be
required. The parameters shown are mostly the same as in
Table~\ref{tab:cutoffpl} but instead of the power law related
parameters, the physical parameters associated with Comptonization and
reflection are shown, namely the electron temperature of the
Comptonizing plasma $kT_\text{e}$ and its optical depth $\tau$, and
the covering factor of the cold reflecting medium
$\Omega/2\pi$. $kT_{\text {in}}$ in this table is either the
temperature of the \texttt{diskbb} component and/or the seed photon
input for the hot plasma.}\label{tab:comptt}
\begin{tabular}{lrrrr}
\hline
\hline
                                    &                Epoch 1&                   Epoch 2&                            Epoch 3&                          Epoch 4\\
\hline
$N_{\text {H}}/10^{22}$ [cm$^{-2}$] & 1.50$^{+0.21}_{-0.26}$&    1.37$^{+0.20}_{-0.15}$&                               1.50&                             1.50\\
$kT_{\text {in}}$ [eV]              &      482$^{+14}_{-16}$&       379$^{+116}_{-378}$&                  441$^{+86}_{-55}$&                501$^{+71}_{-90}$\\
$A_{\text {disk}}/10^{3}$           &    2.5$^{+0.4}_{-0.3}$&                        --&             0.38$^{+0.54}_{-0.04}$&           0.28$^{+0.72}_{-0.07}$\\
$\tau$                              & 0.29$^{+0.43}_{-0.13}$&    0.71$^{+0.16}_{-0.07}$&             1.00$^{+0.21}_{-0.21}$&           0.37$^{+0.24}_{-0.12}$\\
$kT_{\text {e}}$ [keV]              &        64$^{+4}_{-15}$&          78$^{+34}_{-15}$&                    49$^{+29}_{-9}$&                114$^{+32}_{-35}$\\
$E_{\text {Fe}}$ [keV]              & 6.41$^{+0.13}_{-0.24}$&    6.62$^{+0.21}_{-0.30}$&             6.74$^{+0.15}_{-0.26}$&           6.47$^{+0.32}_{-0.25}$\\ 
$EW_{\text {Fe}}$ [eV]              &                  208.0&                      61.0&                               42.6&                             64.3\\ 
$c_{\text {ISGRI}}$                 & 0.76$^{+0.08}_{-0.08}$&    0.83$^{+0.03}_{-0.02}$&             0.82$^{+0.01}_{-0.02}$&           0.87$^{+0.02}_{-0.02}$\\
$c_{\text {SPI}}$                   &                     --&    0.96$^{+0.05}_{-0.04}$&             0.97$^{+0.10}_{-0.09}$&           0.97$^{+0.12}_{-0.11}$\\
$10^{2}$ $\Omega/2\pi$              &                     --&      10.0$^{+5.6}_{-5.6}$&               13.8$^{+5.0}_{-5.5}$&                               --\\
$\chi^2/{\text {dof}}$              &                37.2/34&                   52.5/63&                            57.3/51&                          49.8/52\\
$\chi^2_{\text {red}}$              &                   1.09&                      0.83&                               1.12&                             0.96\\
\hline
\end{tabular}

\end{table}

\subsubsection{The 3--200\,keV Spectrum}\label{sec:fitting}

As a Galactic Center source, \grssp is known to be strongly absorbed
and the $N_\text{H}$ value adopted in most studies is
(1.5$\pm$0.1)$\times10^{22}$\,cm$^{-2}$, as derived by
\citet{mereghetti:97a} from \textsl{ASCA} observations. However,
\citet{keck:01a} report (0.98$\pm$0.08)$\times10^{22}$\,cm$^{-2}$ from
\textsl{ROSAT} observations, \citet{lin:00a} find
(0.93--2.0)$\times10^{22}$\,cm$^{-2}$ from \xtesp observations, and
\citet{goldwurm:01a} determine
(1.74$\pm$0.07)$\times10^{22}$\,cm$^{-2}$ with \textsl{XMM}. Modeling
PCA data starting at 3\,keV, $N_\text{H}$ and the blackbody parameters
are known to be strongly correlated and not well constrained. Here, we
obtain best fits with $N_\text{H}$ values generally well consistent
with the canonical value of 1.5$\times10^{22}$\,cm$^{-2}$ for epochs~1
and 2 \citep[for the epoch~2 \texttt{cutoffpl} fit, though,
$N_\text{H}$ is closer to the lower value of][]{keck:01a}. In the case
of epoch~1 this includes a \texttt{diskbb} component which is
obviously required, while in the case of epoch~2 no thermal component
is needed. For epochs~3 and 4 the best fits with free $N_\text{H}$
result in too low values of (0.03--0.7)$\times10^{22}$\,cm$^{-2}$ and
freezing $N_\text{H}$ to the canonical value does not produce
acceptable fits. Adding a disk blackbody component, however, allows
for good fits with the canonical $N_\text{H}$ (frozen with exception
of the epoch~4 \texttt{cutoffpl} fit). For the Comptonization fit of
epoch~4 this procedure results in a somewhat higher plasma temperature
and lower optical depth compared to the other hard state observations
than without including the disk component
(Table~\ref{tab:comptt}). The same tendency in presence of a disk
blackbody is seen when holding $N_\text{H}$ at the lower value of
\citet{keck:01a}. For all Comptonization fits the blackbody
temperature has been tied to the seed photon temperature of the
\texttt{compTT} component.

Clear residuals in the 6--7\,keV range are present for all epochs when
no iron line is included. The $\chi^2$ values obtained when removing
the Gaussian iron line from the models is given for reference in
Table~\ref{tab:cutoffpl}. Note that the $F$-test may not be used to
test for the presence of a line \citep{protassov:02a}. In epoch~4 the
improvement of the fit when including the iron line is considerably
smaller than for the other epochs and an acceptable fit can be
achieved without the line ($\chi^2_{\text {red}}$=1.09), residuals
remain, however. The iron line is generally narrow, with widths around
or below 0.4\,keV, and consistent with zero. The line energy ranges
from 6.40 to 6.73\,keV and is mostly consistent with 6.4\,keV, i.e.,
neutral Fe. Interestingly, the one exception is epoch~3 where we also
measure the strongest reflection component. A 3--4 times higher line
equivalent width is measured for the soft state epoch~1, consistent
with earlier measurements
\citep{heindl:98a,smith:01a,sidoli:02a}. This mainly reflects the
reduced level of continuum emission during that time, since the line
normalization does not change significantly between the epochs,
including the soft state. It has to be kept in mind, though, that the
Galactic diffuse emission features a strong iron line and that the
iron line parameters obtained from the fits are most likely influenced
by a non-perfect correction for this emission.
  
The parameters we are mainly interested in are those characterizing
the broad band continuum. We caution again that calibration
uncertainties prohibit a statistical comparison with earlier
results. Nevertheless we list earlier results for a qualitative
comparison and to illustrate the overall picture. For epochs~2 to 4 we
find values typical for hard state BHC spectra. For the
phenomenological model the power law indices lie between
1.54$^{+0.01}_{-0.02}$ (epoch~2) and 1.69$^{+0.05}_{-0.05}$ (epoch~4)
and the cutoff energies range from 136$^{+13}_{-16}$\,keV (epoch~3) to
246$^{+26}_{-56}$\,keV (epoch~4). \citet{lin:00a} find
$\Gamma\sim$1.40 (uncertainty of the order of 0.04) and $E_{\text
{cutoff}}\sim$200\,keV (uncertainty of the order of 30\,keV) for their
joint \xtesp and \textsl{CGRO}/OSSE spectrum of 1997 and
\citet{kuznetsov:99a} find $\Gamma=1.0\pm0.3$ and $E_{\text
{cutoff}}=89^{+40}_{-20}$\,keV for their combined 1990--1997
\textsl{GRANAT} data (no cutoff was fit to shorter data sets). With
246\,keV the cutoff energy for epoch~4 is at the limit of what can be
measured with these observations. However, no good fit can be obtained
without cutoff ($\chi^2_{\text {red}}$=2.9).

From the Comptonization models we obtain plasma temperatures of
78$^{+34}_{-15}$\,keV and 49$^{+29}_{-9}$\,keV and optical depths of
0.71$^{+0.16}_{-0.07}$ and 1.00$^{+0.21}_{-0.21}$ for epochs~2 and
3. Here the values of \citet{lin:00a} are $kT_{\text {e}}\sim$52\,keV
(uncertainty of the order of 7\,keV) and $\tau\sim$3.4 (uncertainty of
the order of 0.3). Similarly, \citet{sidoli:02a} find values of
$kT_{\text {e}}=44^{+146}_{-7}$\,keV and $\tau=3.6^{+0.4}_{-2.3}$ for
their \textsl{BeppoSAX} data set. In both of these cases the higher
optical depth is probably mainly due to two effects, first the fact that
no reflection component has been included in the models and second
that the sphere$+$disk geometry has been used. We see a moderate trend
towards higher values of $\tau$ when switching from slab to
sphere$+$disk geometry in our fits.

\citet{kuznetsov:99a} obtain $kT_{\text {e}}=41^{+7}_{-5}$\,keV and
$\tau=1.2\pm0.2$. However, they were using a predecessor to
\texttt{compTT}, namely the model of \citet{sunyaev:80}, therefore
their results cannot be directly compared to ours. Also, these values
reflect the average over a wide range of $kT_{\text {e}}$ and $\tau$
values obtained from their two observation periods each year.

\begin{figure}
\includegraphics[width=88mm]{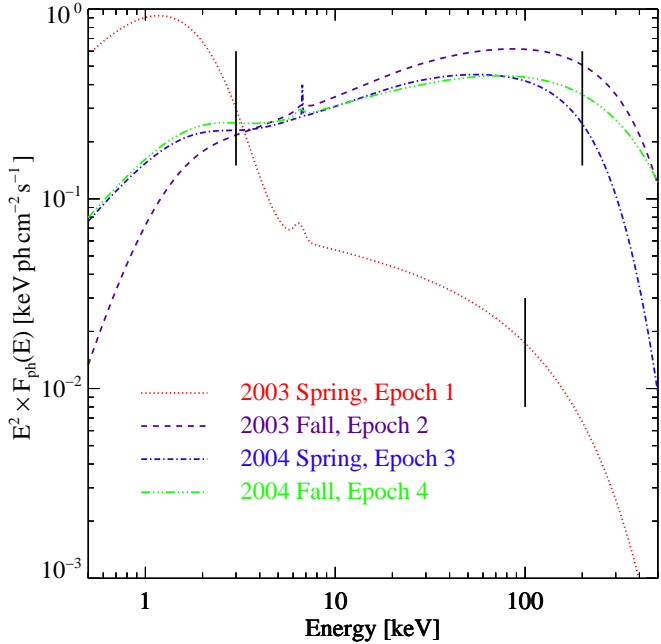}
\caption{Unfolded, unabsorbed model spectra corresponding to the
  \texttt{compTT} fits listed in Table~\ref{tab:comptt}. The $N_{\text
  H}$ values quoted in that table have been used for the flux
  correction. Vertical lines denote the range of the modeled
  data.}\label{fig:modelunabs}
\end{figure}

In the Comptonization fits we also allow for reflection of the
Comptonized radiation of a cold accretion disk and find reflection
factors of 10.0$^{+5.6}_{-5.6}$\% and 13.8$^{+5.0}_{-5.5}$\% for
epochs~2 and 3, respectively. No reflection is detected in
epoch~4. From the \texttt{cutoffpl} fits it is also clear that the
epoch 4 spectrum is less curved than the other two hard state spectra.
With $kT_{\text {e}}=$114$^{+32}_{-35}$\,keV and
$\tau=$0.37$^{+0.24}_{-0.12}$ the Comptonization parameters of epoch~4
correspond to the hottest and most transparent plasma among the hard
state observations. While the latter might be an artifact due to the
introduction of the disk blackbody necessary to constrain $N_{\text
H}$ (the Compton-$y$ changes only slightly)\footnote{Note, however,
that apart from the already quoted effect of obtaining a higher
reflection fraction, the best fit parameters obtained with
\texttt{compPS} for epoch~4 show the same tendency ($kT_{\text
{in}}$=521\,eV, $kT_{e}$=102\,keV, $\tau$=0.63, $\Omega/2\pi$=0.05,
and $\chi_{\text {red}}^2$=0.95) although no additional disk blackbody
has been included.}, a possible physical origin for the differences
observed in the epoch~4 spectrum is suggested by the occurrence of one
of the sudden moderate drops in the PCA rate during this time (see
Fig.\ref{fig:longlc} and Sect.~\ref{sec:evol}).

In general, the range of different results for the hard state
parameters is not too surprising in the light of the considerable
long term variations known to be present even within the hard state
(Fig.~\ref{fig:longlc}), however, it is also clear that \intsp
calibration caveats apply. With $kT_{\text {e}}\sim$50--60\,keV,
$\tau\sim$1.0--1.2, and reflection fractions of 17--24\% the
\texttt{compTT} fits of \citet{pottschmidt:03a} to a set of \intsp and
\xtesp observations of Cyg~X-1 result in similar parameters as
observed for epoch~3.

As expected from the long term evolution of the light curves, the
spectrum of epoch~1 differs considerably from the others. In both
models an additional soft component of comparable strength is clearly
present. We obtain a multicolor disk blackbody temperature of
477$^{+11}_{-27}$\,eV from the power law fit and of
482$^{+14}_{-16}$\,eV from the Comptonization fit. This is consistent
with the 2001 dim soft state where \citet{smith:01a} found a disk
blackbody temperature of 464$\pm$7\,eV with the PCA,
\citet{miller:02a} give values of 340$\pm$10\,eV and 600$\pm$10\,eV,
depending on $N_{\text H}$, for \textsl{XMM} observations, and
\citet{heindl:02a} find 505$\pm$7\,eV with \textsl{Chandra}.  Based on
these previously measured soft state values and since the soft state
spectrum is dominated by disk emission below $\sim$5\,keV, we believe
that the values quoted above give a realistic measure of the
temperature. Not surprisingly the seed photon / disk temperature is
not well constrained in the hard state observations. For epoch~3,
e.g., the disk temperatures obtained from the cutoff power law and the
Comptonization fits are formally not consistent but what is consistent
is the fact that in both cases the disk component is needed if
$N_{\text H}$ is assumed to lie within the range of previously
measured values. Where a disk blackbody component was included in the
hard state fits it never dominates the soft spectrum. With
$\Gamma$=2.29$^{+0.10}_{-0.05}$ the power law is significantly steeper
in epoch~1 but does not quite reach the value of
$\Gamma$=2.75$\pm$0.12 observed in 2001 March \citep{smith:01a}. No
cutoff is detected but during this time the high energy flux was
comparatively weak and the spectrum could only be obtained out to
100\,keV. The steepness of the spectrum translates into a small
optical depth of 0.29$^{+0.43}_{-0.13}$ in the Comptonization fit,
while the temperature of the hot plasma is found to be
64$^{+4}_{-15}$\,keV, i.e., not significantly different from the hard
state epochs~2 and 3.

\section{Discussion}\label{sec:discussion}

Is the dim soft state of \grssp really that different from the soft
states observed in other sources? In Fig.~\ref{fig:modelunabs} the
unfolded, unabsorbed model spectra corresponding to the
\texttt{compTT} fits are shown. The typical pivoting between the soft
state spectrum and the three hard state spectra is seen. With
$\sim$3\,keV the pivot energy lies considerably lower than for Cyg
X-1, where a value of about $\sim$10\,keV is observed
\citep{zdz:02a,wilms:05a}. However, taking the nature of \grssp as low
mass X-ray binary (LMXB) and Roche lobe accretor into account, its
behavior might be more akin to the state transitions displayed by LMXB
BHC transients than to those of the high mass X-ray binary (HMXB) and
focused wind accretor \mbox{Cyg X-1}, i.e., hysteresis might play an
important role. In the following the term ``hysteresis'' is used to
describe the existence of an ``overlap region'' in luminosity in which
both, soft and hard states can occur \citep[see,
e.g.,][]{miyamoto:95a,zdz:04b,meyer:05a}\footnote{The flux
derivative/hardness correlation may represent the extension of this
hysteretic behavior into the hard state \citep{smith:01b}.}. According
to the rough estimate for the bolometric luminosity that can be
derived from our fits (using a distance estimate of 8.5\,kpc based on
the assumption of a near-GC location of \grs), the 2003 dim soft state
is 0--20\% less luminous than the hard state, depending on the hard
state epoch and spectral model used for comparison. For a
10\,$M_{\sun}$ black hole the hard state luminosities that we measure
correspond to 2--3\% $L_{\text{Edd}}$. The differences between the
states in terms of fluxes in different energy bands have been
presented in Table~\ref{tab:fluxes}.

Before comparing the range of hysteretic behavior observed in \grssp
to other sources, we note that another possible reason for observing
reduced soft state luminosities might be a geometric effect introduced
by the inclination $i$ of the system: In the hard state the
geometrically thick hot plasma is present which can be assumed to
radiate approximately isotropically. In the soft state only the
decaying accretion disk remains which is geometrically thin with a
luminosity $\propto \cos i$ \citep{frank:92}. If the system is viewed
close to edge-on the projected area of the inner disk is comparatively
small, allowing only for a small percentage of the disk luminosity to
reach the observer. In addition, X-rays from the inner disk may be
further obscured due to flaring of the outer disk \citep{narayan:05a}.

\begin{figure}
\includegraphics[width=88mm]{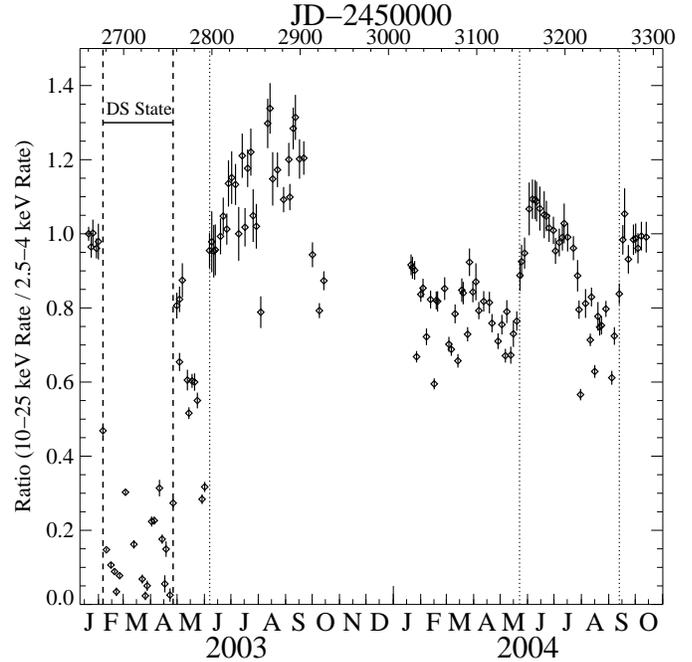}
\caption{Ratio of the 10--25\,keV and 2.5--4\,keV PCA count
  rates. Dashed and dotted vertical lines denote the dim soft state of
  2003 and times of sudden hardening, respectively.}\label{fig:ratio}
\end{figure}

GX~339$-$4 is the source for which hysteresis in the above sense has
been best studied so far. Depending on the energy range, its lowest
soft state flux can lie a factor of 2.5--10 below the brightest hard
state flux \citep{nowak:01b,zdz:04a,belloni:05a}. \citet{nowak:01b},
e.g., find that the 3--9\,keV soft state flux can be less than half
the hard state flux, similar to what we find for \grssp
(Table~\ref{tab:fluxes}). The bolometric flux of GX~339$-$4 in the
soft state can be up to an order of magnitude lower than in the hard
state \citep{zdz:04a}, an even more extreme behavior than indicated by
our bolometric estimates for \grs. Accordingly, the schematic picture
which has recently been developed of the ``\textsf{q}-shaped'' tracks
followed by black hole transients in the hardness-intensity diagram
over an outburst includes a large range of soft state intensities
\citep[$\gtrsim {\cal O}(1)$,][]{fender:04a}, not necessarily
exceeding the highest hard state ones. Although we concentrate on
average soft state parameters in this work, we want to note that the
hardness-intensity diagrams for the 2003 soft state that can be
derived from the energy-resolved PCA light curves show a
counterclockwise evolving pattern comparable to transients. Due to the
pronounced soft state it is rather slightly ``\textsf{p}-shaped'' but
otherwise qualitatively very similar (a detailed quantitative
comparison is beyond the scope of this work). Overall it seems that
\grs's dim soft state of 2003 -- and also the even dimmer one of 2001
-- are no remarkable states for a non high mass BHC \citep[see
also][]{remillard:05a}. This is especially true if part of the
luminosity reduction in the dim soft state is due to the inclination
effect described above. How about the overall state evolution, though?
Can the occurrence of the dim soft states be understood in the frame
of the outburst evolution scheme mentioned above? In the following two
paragraphs we discuss this question on the basis of the light curves
displayed in Figs.~\ref{fig:lcpanel} and~\ref{fig:longlc}. Note,
however, that a detailed spectral analysis of the individual PCA
pointings is beyond the scope of this paper.

The initial phase of the state change consists of a sudden drop of the
$>4$\,keV count rate around JD 2452680 (2003 February) and a
simultaneous moderate brightening of the thermal component, observed
as a $\sim$70$\%$ increase in the 2.5--4\,keV count rate. Only two
monitoring observations find the source in this phase, i.e., it lasted
roughly a week. During the following weeks the dim soft state is
observed: The hard emission does not recover until end of 2003 April
and after the initial increase the soft emission decays slowly to a
low hard state level. The initial outburst-like phase is similar to a
canonical transition to a soft state but with the system not settling
into a state with a stable thermal component. In this sense the
episode is a ``failed state transition''. The short soft flare may
reflect an actual change in the accretion disk parameters (e.g., a
temperature change and/or a change of the inner disk
radius). Alternatively, the increase in soft photon flux could at
least partly be caused by the disappearance of the Comptonizing
medium, i.e., the soft photons acting as seed photons in the hard
state are now emerging without being Comptonized. While the 2001 dim
soft state showed no initial flaring of the 2.5--4\,keV count rate, a
soft excess compared to the hard state level also became visible in
the unabsorbed spectrum.

In contrast to the weeks long soft X-ray flares of \mbox{Cyg X-1},
however, for which the term ``failed state transition'' was coined
\citep{pottschmidt:00a,pottschmidt:03a}, \grssp does not settle back
into the hard state after the flaring but the hard component stays
``off''. In the case of the 2001 dim soft state \citet{smith:01a}
suggested a sudden shutoff of mass transfer from the companion being
responsible for the ``off'' phase. Put into the context of the ``q''
pattern of transient outbursts, the dim soft states of \grssp could
therefore well represent the thermally dominated outburst phase since
the main decay track proceeds through this state
\citep{remillard:05a}. The hard state of \grs, also covering a
considerable range of luminosities, would then correspond to phases of
rising and peak luminosities, again consistent with transient
outbursts. As mentioned in Sect.~\ref{sec:intro}, additional
intermediate states -- or failed state transitions -- of \grssp have
been observed \citep{mereghetti:94a,goldwurm:01a,heindl:02b}, further
completing the outburst picture.

Another interesting property of the dim soft state is the fact that
the decay of the hard and soft spectral components is governed by two
different time scales. This has been studied in detail for the 2001
dim soft state by \citet{smith:01a} who found that while the power law
flux decreased by an order of magnitude from one monitoring
observation to the next, the disk black body flux decayed on a time
scale of $\sim$28\,d. This behavior is also visible in the 2003 light
curves (Fig.~\ref{fig:lcpanel}), especially in the fast decline of the
10--25\,keV rates and the much slower trend in the 2.5--4\,keV rates
after the initial drop down from the ``failed state transition''
level. The source also shows several drops of the hard component for
durations of only a few days or less, e.g., around JD 2451932
\citep[2001 January, see Fig.~\ref{fig:longlc} and][]{smith:01a},
2452364 (2002 March, Fig.~\ref{fig:longlc}), 2452788 (2003 May,
Fig.~\ref{fig:lcpanel}), or 2452855 (2003 August,
Fig.~\ref{fig:lcpanel}). All these quasi-independent changes of the
hard and soft spectral component further support the interpretation of
the behavior of \grssp in terms of two different accretion flows. As
shown by \citet{smith:01a}, the model of \citet{chakrabarti:95a} can
explain many of the observations. As already mentioned, this model
assumes that proportional accretion rate changes introduced to both
flows at large radii propagate with nearly the free-fall time scale
through the Comptonizing medium and independently on the viscous time
scale through the accretion disk. Different propagation speeds are a
general feature of the model, i.e., they are not restricted to its
high accretion rate soft state associated with bulk motion
Comptonization. For lower accretion rates complicated dependencies of
spectral hardness and accretion rate are possible, covering the
correlation between the flux derivative and the spectral hardness as
well as the dim soft state \citep{chakrabarti:95a,smith:01a}. The
strength of these time delay effects increases for larger accretion
disks and there are indications that such a picture might be generally
applicable for Roche lobe overflow transients: a state transition due
to a sudden change in the power law component during a time when the
disk black body parameters evolved smoothly has recently also been
seen in the black hole transient \object{H1743$-$322}, in this case
marking the transition between the thermally dominant and the
intermediate state \citep{kalemci:05b}.

In addition to soft episodes we also observe several occurrences of a
rather sudden hardening (Fig.~\ref{fig:ratio}) mainly due to declines
of the soft component, visible, e.g., in the 2.5--4\,keV rates around
JD 2452797 (2003 June), 2453148 (2004 May), or 2453261 (2004
September). In the first case this is clearly related to a preceding
drop of the hard component. In the second case the drop happens at the
end of a months long decline of the count rates in all energy bands
which is especially visible in the \intsp range (epoch~3). The
situation is less clear for the third occurrence (during epoch~4) but
the 10--25\,keV light curve also indicates a preceding decline. Due to
this overall picture and the probably affected broad band spectrum of
epoch~4, we consider it less likely that the hard episodes are caused,
e.g., by absorption events, but are rather another example of the
quasi-independent behavior of the hard and soft component on different
time scales. Interestingly, a similar episode of sudden hardening has
also been observed for the ``two-flow source'' 1E~1740.7$-$2942
\citep{smith:01b}.

Finally, while the hard state parameters have been discussed in
Sect.~\ref{sec:fitting} already, we emphasize again that apart from
small peculiarities which might be caused by spectral variations
within the epochs (epoch~4) the epoch-summed hard state spectra can be
well described by cutoff power law and thermal Comptonization
parameters which are compatible with canonical values found for BHCs
in the hard state, e.g., Cyg~X-1.
 
\section{Summary}\label{sec:conclusions}
We have presented analyses of \intsp and \xtesp monitoring
observations of the Galactic Center BHC \grssp obtained in 2003 and
2004. Energy-resolved light curves from 2.5 to 200\,keV have been
studied and broad band spectra accumulated over four $\sim$2--3 months
long \intsp observing epochs have been modeled phenomenologically and
with thermal Comptonization. The main results of this work can be
summarized as follows:
\begin{itemize}
\item From 2003 February to April \grssp entered another dim soft
  state (partly covered by epoch~1) similar to but less prolonged than
  the one observed in 2001.
\item Phenomenological models (dominated by a power law in the dim
  soft state and an exponentially cutoff power law in the hard states)
  as well as thermal Comptonization models (\texttt{compTT} and
  \texttt{compPS}) allow for a good description of the epoch-summed
  spectra.
\item The fit parameters obtained in the hard state are canonical BHC
  hard state parameters, similar to those obtained for Cyg~X-1 and
  generally consistent with previous results obtained for \grs. The
  hard state in \grssp is known to be intrinsically variable. Since it
  is not clear how much of the range of the fit parameters observed
  between the hard state epochs is due to calibration uncertainties,
  however, these differences are not interpreted further.
\item In the dim soft state the flux is only higher than in the hard
  state below 3--4\,keV. In all other energy bands the flux is
  considerably lower. A tentative estimate of the bolometric
  luminosity, however, shows a reduction of only 0--20\% compared to
  the hard state epochs.
\item While the dim soft state is different from the soft state in
  persistent HMXBs like Cyg~X-1 or \object{LMC~X-3} where softening is
  associated with higher luminosities (mass accretion rates), it is
  well within the range of hysteretic behavior displayed by LMXB
  transients like GX~339$-$4, where a large range of soft state
  intensities, not necessarily exceeding the highest hard state ones,
  is observed (``q'' pattern in the hardness-intensity diagram). The
  dim soft state would thus correspond to the outburst decay of a
  transient. This can be understood since \grssp most likely has a low
  mass companion and is accreting via Roche lobe overflow.
\item As discovered for the 2001 dim soft state by \citet{smith:01a},
  the decay of the soft and hard component progresses on different
  time scales which can be understood if the accretion flows through
  the cool disk and the hot plasma are independent in the sense of the
  two-flow model of \citet{chakrabarti:95a}. The evolution of the
  light curves, especially comparing the 2.5--4\,keV and 10--25\,keV
  PCA bands during the 2003 dim soft state also suggests different
  time scales for changes in the two spectral components.
\item In addition, several instances are observed during which
  predominantly one of the two spectral components shows a substantial
  flux change from one monitoring observation to the next, further
  supporting the picture of at least partly independent flows.
\end{itemize}

\begin{acknowledgements}

We thank J\"orn Wilms for helpful discussions. This work has been
partly funded by NASA contract NAS5-30720 (KP) as well as by NASA
grants NAG5-13576 and NNG04GP41G (DMS, NB). AAZ and PL have been
supported by KBN grants 1P03D01827, 1P03D01727, 1P03D01128,
PBZ-KBN-054/P03/2001 and 4T12E04727. This work is based on
observations with \int, an ESA project with instruments
and science data centre funded by ESA member states (especially the PI
countries: Denmark, France, Germany, Italy, Switzerland, Spain), Czech
Republic and Poland, and with the participation of Russia and the USA.
We thank the \xtesp schedulers for making the years long
monitoring campaign of \grssp possible. KP thanks the Aspen Center for
Physics for its hospitality during the final stages of the preparation
of this paper.

\end{acknowledgements}


\begin{thebibliography}{}

\bibitem[\protect\astroncite{Belloni et~al.}{2005}]{belloni:05a}
Belloni T., Homan J., Casella P., et~al., 2005, A\&A 440, 207

\bibitem[\protect\astroncite{{Chakrabarti} \&
  {Titarchuk}}{1995}]{chakrabarti:95a}
{Chakrabarti} S., {Titarchuk} L.G.,  1995, ApJ 455, 623

\bibitem[\protect\astroncite{{Coppi}}{1999}]{coppi:99a}
{Coppi} P.S.,  1999,
\newblock in High Energy Processes in Accreting Black Holes,
ed. Poutanen J.\& Svensson R. (San Francisco: Astron.\ Soc.\ Pacific),
ASP Conf.\ Ser. 161, 375  

\bibitem[\protect\astroncite{{Dubath} et~al.}{2005}]{dubath:05a}
{Dubath} P., {Kn{\"o}dlseder} J., {Skinner} G.K., et~al., 2005, MNRAS 357, 420

\bibitem[\protect\astroncite{{Eikenberry} et~al.}{2001}]{eikenberry:01a}
{Eikenberry} S.S., {Fischer} W.J., {Egami} E., {Djorgovski} S.G.,  2001, ApJ
  556, 1

\bibitem[\protect\astroncite{{Fender} et~al.}{2004}]{fender:04a}
{Fender} R.P., {Belloni} T.M., {Gallo} E.,  2004, MNRAS 355, 1105

\bibitem[\protect\astroncite{Frank et~al.}{1992}]{frank:92}
Frank J., King A., Raine D.,  1992,
\newblock Accretion Power in Astrophysics (Cambridge: Univ.\ Cambridge Press)

\bibitem[\protect\astroncite{{Gilfanov} et~al.}{1993}]{gilfanov:93a}
{Gilfanov} M., {Churazov} E., {Sunyaev} R., et~al., 1993, ApJ 418, 844

\bibitem[\protect\astroncite{{Goldwurm} et~al.}{2003}]{goldwurm:03a}
{Goldwurm} A., {David} P., {Foschini} L., et~al., 2003, A\&A 411, L223

\bibitem[\protect\astroncite{Goldwurm et~al.}{2001}]{goldwurm:01a}
Goldwurm A., Israel D., Goldoni P., et~al., 2001,
\newblock in Proc.\ of the Gamma-Ray Astrophysics 2001 Symposium
(Woodbury: AIP), AIP Conf.\ Proc. 587, 61

\bibitem[\protect\astroncite{{Grebenev} et~al.}{1997}]{grebenev:97a}
{Grebenev} S.A., {Pavlinsky} M.N., {Sunyaev} R.A.,  1997,
\newblock in The Transparent Universe, ed. Winkler C., Courvoisier
T.J.L. \& Durouchoux P. (Noordwijk: ESA Publications Division), ESA SP
382, 183 

\bibitem[\protect\astroncite{{Heindl} \& {Smith}}{1998}]{heindl:98a}
{Heindl} W.A., {Smith} D.M.,  1998, ApJ 506, L35

\bibitem[\protect\astroncite{{Heindl} \& {Smith}}{2002a}]{heindl:02b}
{Heindl} W.A., {Smith} D.M.,  2002a, ApJ 578, L125

\bibitem[\protect\astroncite{{Heindl} \& {Smith}}{2002b}]{heindl:02a}
{Heindl} W.A., {Smith} D.M.,  2002b,
\newblock in The High Energy Universe at Sharp Focus: Chandra Science,
ed. Schlegel E.M. \& Vrtilek S.D. (San  Francisco: Astron.\ Soc. of the
Pacific), ASP Conf.~Series 262, 241 

\bibitem[\protect\astroncite{Jahoda et~al.}{1996}]{jahoda:96}
Jahoda K., Swank J.H., Giles A.B., et~al., 1996,
\newblock in {EUV}, X-Ray, and Gamma-Ray Instrumentation for Astronomy
	  {VII}, ed. Siegmund O.H. (Bellingham, WA: SPIE), Proc.\ SPIE
	  2808, 59  

\bibitem[\protect\astroncite{{Kalemci} et~al.}{2006}]{kalemci:05b}
{Kalemci} E., {Tomsick} J.A., {Rothschild} R.E., et~al., 2006, ApJ, 639, 340

\bibitem[\protect\astroncite{{Keck} et~al.}{2001}]{keck:01a}
{Keck} J.W., {Craig} W.W., {Hailey} C.J., et~al., 2001, ApJ 563, 301

\bibitem[\protect\astroncite{{Kuznetsov} et~al.}{1999}]{kuznetsov:99a}
{Kuznetsov} S.I., {Gilfanov} M.R., {Churazov} E.M., et~al., 1999, Astron.\
  Let.\ 25, 351

\bibitem[\protect\astroncite{{Lebrun} et~al.}{2003}]{lebrun:03a}
{Lebrun} F., {Leray} J.P., {Lavocat} P., et~al., 2003, A\&A 411, L141

\bibitem[\protect\astroncite{{Lin} et~al.}{2000}]{lin:00a}
{Lin} D., {Smith} I.A., {Liang} E.P., et~al., 2000, ApJ 532, 548

\bibitem[\protect\astroncite{{Lund} et~al.}{2003}]{lund:03a}
{Lund} N., {Budtz-J{\o}rgensen} C., {Westergaard} N.J., et~al., 2003, A\&A 411,
  L231

\bibitem[\protect\astroncite{{Magdziarz} \& {Zdziarski}}{1995}]{magdziarz:95a}
{Magdziarz} P., {Zdziarski} A.A.,  1995, MNRAS 273, 837

\bibitem[\protect\astroncite{{Main} et~al.}{1999}]{main:99a}
{Main} D.S., {Smith} D.M., {Heindl} W.A., et~al., 1999, ApJ 525, 901

\bibitem[\protect\astroncite{{Mandrou}}{1990}]{mandrou:90a}
{Mandrou} P.,  1990, IAU Circ. 5032

\bibitem[\protect\astroncite{{Marti} et~al.}{1998}]{marti:98a}
{Marti} J., {Mereghetti} S., {Chaty} S., et~al., 1998, A\&A 338, L95

\bibitem[\protect\astroncite{{Mereghetti} et~al.}{1994}]{mereghetti:94a}
{Mereghetti} S., {Belloni} T., {Goldwurm} A.,  1994, ApJ 433, L21

\bibitem[\protect\astroncite{{Mereghetti} et~al.}{1997}]{mereghetti:97a}
{Mereghetti} S., {Cremonesi} D.I., {Haardt} F., et~al., 1997, ApJ 476, 829

\bibitem[\protect\astroncite{{Meyer-Hofmeister} et~al.}{2005}]{meyer:05a}
{Meyer-Hofmeister} E., {Liu} B.F., {Meyer} F.,  2005, A\&A 432, 181

\bibitem[\protect\astroncite{{Miller} et~al.}{2002}]{miller:02a}
{Miller} J.M., {Wijnands} R., {Rodriguez-Pascual} P.M., et~al., 2002, ApJ 566,
  358

\bibitem[\protect\astroncite{{Miyamoto} et~al.}{1995}]{miyamoto:95a}
{Miyamoto} S., {Kitamoto} S., {Hayashida} K., {Egoshi} W.,  1995, ApJ 442, L13

\bibitem[\protect\astroncite{{Narayan} \& {McClintock}}{2005}]{narayan:05a}
{Narayan} R., {McClintock} J.E.,  2005, ApJ 623, 1017

\bibitem[\protect\astroncite{Nowak et~al.}{2002}]{nowak:01b}
Nowak M.A., Wilms J., Dove J.B.,  2002, MNRAS 332, 856

\bibitem[\protect\astroncite{Pottschmidt et~al.}{2003}]{pottschmidt:03a}
Pottschmidt K., Wilms J., Chernyakova M., et~al., 2003, A\&A 411, L383

\bibitem[\protect\astroncite{Pottschmidt et~al.}{2004}]{pottschmidt:04a}
Pottschmidt K., Wilms J., Nowak M.A., et~al., 2004,
\newblock in Proc.\ 5th {INTEGRAL} Workshop: The INTEGRAL Universe, ed.
Sch\"onfelder V., Lichti G., \& Winkler C. (Noordwijk: ESA). ESA
SP-552, 345 

\bibitem[\protect\astroncite{Pottschmidt et~al.}{2000}]{pottschmidt:00a}
Pottschmidt K., Wilms J., Nowak M.A., et~al., 2000, A\&A 357, L17

\bibitem[\protect\astroncite{{Poutanen} \& {Svensson}}{1996}]{poutanen:96a}
{Poutanen} J., {Svensson} R.,  1996, ApJ 470, 249

\bibitem[\protect\astroncite{{Protassov} et~al.}{2002}]{protassov:02a}
{Protassov} R., {van Dyk} D.A., {Connors} A., et~al., 2002, ApJ 571, 545

\bibitem[\protect\astroncite{Remillard}{2005}]{remillard:05a}
Remillard R.A.,  2005,
\newblock in {Texas@Stanford 2004}, ed. Chen P., SLAC Electronic Conference
  Proceedings Archive, astro-ph/0504129

\bibitem[\protect\astroncite{{Rodriguez} et~al.}{1992}]{rodriguez:92a}
{Rodriguez} L.F., {Mirabel} I.F., {Marti} J.,  1992, ApJ 401, L15

\bibitem[\protect\astroncite{{Rothstein} et~al.}{2002}]{rothstein:02a}
{Rothstein} D.M., {Eikenberry} S.S., {Chatterjee} S., et~al., 2002, ApJ 580,
  L61

\bibitem[\protect\astroncite{{Sidoli} \& {Mereghetti}}{2002}]{sidoli:02a}
{Sidoli} L., {Mereghetti} S.,  2002, A\&A 388, 293

\bibitem[\protect\astroncite{{Skinner} \& {Connell}}{2003}]{skinner:03a}
{Skinner} G., {Connell} P.,  2003, A\&A 411, L123

\bibitem[\protect\astroncite{{Smith} et~al.}{2001a}]{smith:01a}
{Smith} D.M., {Heindl} W.A., {Markwardt} C.B., {Swank} J.H.,  2001a, ApJ 554,
  L41

\bibitem[\protect\astroncite{{Smith} et~al.}{1997}]{smith:97a}
{Smith} D.M., {Heindl} W.A., {Swank} J., et~al., 1997, ApJ 489, L51

\bibitem[\protect\astroncite{{Smith} et~al.}{2002a}]{smith:02a}
{Smith} D.M., {Heindl} W.A., {Swank} J.H.,  2002a, ApJ 578, L129

\bibitem[\protect\astroncite{{Smith} et~al.}{2002b}]{smith:01b}
{Smith} D.M., {Heindl} W.A., {Swank} J.H.,  2002b, ApJ 569, 362

\bibitem[\protect\astroncite{{Smith} et~al.}{2001b}]{smith:01c}
{Smith} D.M., {Heindl} W.A., {Swank} J.H., {Markwardt} C.B.,  2001b, ATEL 66

\bibitem[\protect\astroncite{{Strong} et~al.}{2003}]{strong:03a}
{Strong} A.W., {Bouchet} L., {Diehl} R., et~al., 2003, A\&A 411, L447

\bibitem[\protect\astroncite{{Sunyaev} et~al.}{1991}]{sunyaev:91a}
{Sunyaev} R., {Churazov} E., {Gilfanov} M., et~al., 1991, A\&A 247, L29

\bibitem[\protect\astroncite{Sunyaev \& Titarchuk}{1980}]{sunyaev:80}
Sunyaev R.A., Titarchuk L.G.,  1980, A\&A 86, 121

\bibitem[\protect\astroncite{{Terrier} et~al.}{2003}]{terrier:03a}
{Terrier} R., {Lebrun} F., {Bazzano} A., et~al., 2003, A\&A 411, L167

\bibitem[\protect\astroncite{Titarchuk}{1994}]{tit:94}
Titarchuk L.,  1994, ApJ 434, 570

\bibitem[\protect\astroncite{{Vedrenne} et~al.}{2003}]{vedrenne:03a}
{Vedrenne} G., {Roques} J.P., {Sch{\" o}nfelder} V., et~al., 2003, A\&A 411,
  L63

\bibitem[\protect\astroncite{Wilms et~al.}{2006}]{wilms:05a}
Wilms J., Nowak M.N., Pottschmidt K., et~al., 2006, A\&A 447, 245

\bibitem[\protect\astroncite{{Zdziarski} \& {Gierli{\' n}ski}}{2004}]{zdz:04b}
{Zdziarski} A.A., {Gierli{\' n}ski} M.,  2004, Progress of Theoretical Physics
  Supplement 155, 99

\bibitem[\protect\astroncite{{Zdziarski} et~al.}{2004}]{zdz:04a}
{Zdziarski} A.A., {Gierli{\' n}ski} M., {Miko{\l}ajewska} J., et~al., 2004,
  MNRAS 351, 791

\bibitem[\protect\astroncite{Zdziarski et~al.}{2002}]{zdz:02a}
Zdziarski A.A., Poutanen J., Paciesas W.S., Wen L.,  2002, ApJ 578, 357

\bibitem[\protect\astroncite{Zhang et~al.}{1997}]{zhang:97a}
Zhang S.N., Cui W., Harmon B.A., et~al., 1997, ApJ 477, L95

\end{thebibliography}
\end{document}